\documentclass[twocolumn,showpacs,superscriptaddress,amsmath,amssymb,prc,preprintnumbers]{revtex4-1}
\usepackage[pdftex]{graphicx}
\usepackage{dcolumn}
\usepackage{bm}
\usepackage{ifpdf}
\usepackage{epstopdf}
\usepackage{slashed}
\usepackage{amsfonts}
\usepackage{mathrsfs}
\usepackage{amssymb}
\usepackage{multirow}
\usepackage{CJK}
\usepackage[pdftex]{xcolor}
\allowdisplaybreaks[4]
\begin{document}
\begin{CJK*}{UTF8}{}

  \preprint{RIKEN-QHP-490}
  \preprint{RIKEN-iTHEMS-Report-21}
  \title{Exploring effects of tensor force and its strength via neutron drops}
  \author{Zhiheng Wang (\CJKfamily{gbsn}{王之恒})}
  \affiliation{School of Nuclear Science and Technology, Lanzhou University, Lanzhou 730000, China}

  \author{Tomoya Naito (\CJKfamily{gbsn}{内藤智也})}
  \affiliation{Department of Physics, Graduate School of Science, The University of Tokyo,
    Tokyo 113-0033, Japan}
  \affiliation{RIKEN Nishina Center, Wako 351-0198, Japan}

  \author{Haozhao Liang (\CJKfamily{gbsn}{梁豪兆})}
  \affiliation{Department of Physics, Graduate School of Science, The University of Tokyo,
    Tokyo 113-0033, Japan}
  \affiliation{RIKEN Nishina Center, Wako 351-0198, Japan}

    \author{Wen Hui Long (\CJKfamily{gbsn}{龙文辉})}
  \affiliation{School of Nuclear Science and Technology, Lanzhou University, Lanzhou 730000, China}
  \affiliation{Joint department for nuclear physics, Lanzhou University and Institute of Modern Physics,
    Chinese Academy of Sciences, Lanzhou 730000, China}

  \date{\today}
  \begin{abstract}
    The tensor-force effects on the evolution of spin-orbit splittings in neutron drops are investigated within the framework of the relativistic Hartree-Fock theory. For a fair comparison on the pure mean-field level, the results of the relativistic Brueckner-Hartree-Fock calculation with the Bonn A interaction are adopted as meta-data. Through a quantitative analysis, we certify that the $\pi$-pseudovector ($\pi$-PV) coupling affects the evolutionary trend through the embedded tensor force. The strength of the tensor force is explored by enlarging the strength $f_{\pi}$ of the $\pi$-PV coupling. It is found that weakening the density dependence of $f_{\pi}$ is slightly better than enlarging it with a factor. We thus provide a semiquantitative support for the \textit{renormalization persistency} of the tensor force within the framework of density functional theory. This will serve as important guidance for further development of relativistic effective interactions with particular focus on the tensor force.
  \end{abstract}
  \maketitle
\end{CJK*}

    \section{Introduction}
  \par
  The nuclear tensor force is one of the most important components of the bare nucleon-nucleon interaction~\cite{
    Fayache1997Phys.Rep.290.201,
    Sagawa2014Prog.Part.Nucl.Phys.76.76}.
  In recent decades, the effects of the tensor force in the nuclear mediums have also been intensively investigated, after being neglected for a long time.
  It has been shown that the tensor force plays a crucial role in shell-structure evolution~\cite{
    Nakada2003Phys.Rev.C68.014316,
    Otsuka2005Phys.Rev.Lett.95.232502,
    Brown2006Phys.Rev.C74.061303,
    Otsuka2006Phys.Rev.Lett.97.162501,
    Brink2007Phys.Rev.C75.064311,
    Lesinski2007Phys.Rev.C76.014312,
    Long2008Europhys.Lett.82.12001,
    Zou2008Phys.Rev.C77.014314,
    Stanoiu2008Phys.Rev.C78.034315,
    Nakada2008Phys.Rev.C78.054301,
    Moreno-Torres2010Phys.Rev.C81.064327,
    Kaneko2011Phys.Rev.C83.014320,
    Wang2011Phys.Rev.C83.054305,
    Wang2011Phys.Rev.C84.044333,
    Dong2011Phys.Rev.C84.014303,
    Anguiano2012Phys.Rev.C86.054302,
    Tsunoda2014Phys.Rev.C89.031301,
    Shi2017Phys.Rev.C95.034307,
    Shen2019Phys.Rev.C99.034322,
    Chen2019Phys.Rev.Lett.122.212502,
    Wang2019Chin.Phys.C43.124106,
    Dong2020Phys.Rev.C101.014305,
    Nakada2020Int.J.Mod.Phys.E29.1930008},
  spin-isospin excitations~\cite{
    Bai2009Phys.Lett.B675.28,
    Bai2009Phys.Rev.C79.041301,
    Bai2010Phys.Rev.Lett.105.072501,
    Bai2011Phys.Rev.C83.054316,
    Bai2011Phys.Rev.C84.044329,
    Minato2013Phys.Rev.Lett.110.122501},
  and giant resonances~\cite{
    Cao2009Phys.Rev.C80.064304,
    Cao2011Phys.Rev.C83.034324,
    Wen2014Phys.Rev.C89.044311}.
  In spite of such achievements, open questions remain about the properties of the in-medium tensor force, i.e., the effective tensor force.
  Among the most intricate challenges is the problem surrounding the constraint of the strength of the tensor force~\cite{Sagawa2014Prog.Part.Nucl.Phys.76.76}.
  To achieve this goal, one needs to determine the observables that are sensitive to the effective tensor force~\cite{
    Lesinski2007Phys.Rev.C76.014312,
    Anguiano2011Phys.Rev.C83.064306,
    Sagawa2014Prog.Part.Nucl.Phys.76.76,
    Wang2019Chin.Phys.C43.114101,
    Su2019Chin.Phys.C43.064109}.
  By fitting to these observables based on certain many-body theories, one can expect to pin down the sign and strength of the tensor force.
  \par
  The nuclear density functional theory (DFT)~\cite{
    Bender2003Rev.Mod.Phys.75.121,
    Vretenar2005Phys.Rep.409.101,
    Meng2006Prog.Part.Nucl.Phys.57.470,
    Niksic2011Prog.Part.Nucl.Phys.66.519,
    Nakatsukasa2016Rev.Mod.Phys.88.045004,
    Meng2016}
  is currently the only candidate that can be applied to almost the whole nuclear chart, except for very light nuclei.
  Within both nonrelativistic and relativistic DFT, the parameters of the effective interactions are usually determined by fitting to the bulk nuclear properties, such as the masses and radii of the finite nuclei, as well as the empirical knowledge of infinite nuclear matter.
  However, these bulk properties are found to be, in general, not sensitive to the tensor components in effective interactions.
  In particular, by adding the Fock term of the pion exchange, which is one of the most important carriers of the tensor force in the relativistic nuclear forces, on top of the conventional relativistic mean-field (RMF) theory, Lalazissis \textit{et~al.}~\cite{Lalazissis2009PRC80.041301(R)}
  found that the bulk properties of spherical finite nuclei and infinite nuclear matter disfavor the tensor force, i.e., the optimal fit is achieved for the vanishing pion field.
  \par
  Moreover, the tensor force has the characteristic property of spin dependence. It can significantly affect the shell structure of nuclei, especially those located far away from the stability line~\cite{Otsuka2005Phys.Rev.Lett.95.232502}.
  One of the most famous benchmarks is the evolution of the energy difference between the proton states $1h_{11/2}$ and $1g_{7/2}$ in the $\mathrm{Sn}$ ($Z = 50$) isotopes and that between the neutron states $1i_{13/2}$ and $1h_{9/2}$ in the $N = 82$ isotones~\cite{Schiffer2004Phys.Rev.Lett.92.162501}.
  Based on the Skyrme Hartree-Fock (SHF)~\cite{Colo2007Phys.Lett.B646.227}, Gogny Hartree-Fock (GHF)~\cite{Otsuka2006Phys.Rev.Lett.97.162501}, and the relativistic Hartree-Fock (RHF)~\cite{
  Long2008Europhys.Lett.82.12001} theories, it was found that the tensor force plays a crucial role in reproducing the empirical trend of the shell structure mentioned above.
  Accordingly, by reproducing the shell-structure evolution, one can expect to calibrate the strength of the effective tensor force.
  \par
  Nevertheless, the single-particle states observed experimentally are usually fragmented~\cite{
    Bertsch1983Rev.Mod.Phys.55.287,
    Sorlin2008Prog.Part.Nucl.Phys.61.602,
    Kay2008Phys.Lett.B658.216,
    Kay2011Phys.Rev.C84.024325}
  due to, for example, the coupling with low-lying vibrations, which is related to the quenching of the spectroscopic factors.
  The distraction arising from the beyond-mean-field correlations makes it ambiguous to directly compare the single-particle energies calculated by DFT with the corresponding experimental data.
  A possible solution to eliminate this kind of distraction is to take into account the particle-vibration coupling (PVC) in the theoretical calculations~\cite{
    Litvinova2006PhysRevC.73.044328,
    Colo2010Phys.Rev.C82.064307,
    Afanasjev2015Phys.Rev.C92.044317,
    Karakatsanis2017Phys.Rev.C95.034318}.
  By doing so, the descriptions of the energies, as well as the wave functions, can be improved, although the fragmentation of the single-particle states may remain a problem.
  \par
  Another option to avoid the distraction of the beyond-mean-field correlations is to seek for the \textit{ab initio} calculations which can serve the meta-data,
  instead of the experimental data.
  In the last decade, the \textit{ab initio} calculations have progressed greatly~\cite{
    Carlson2015Rev.Mod.Phys.87.1067,
    Laehde2014Phys.Lett.B732.110,
    Bogner2013Comput.Phys.Commun.184.2235,
    Jansen2014Phys.Rev.Lett.113.142502}.
  In particular, Shen \textit{et al.}~have established the self-consistent relativistic Brueckner-Hartree-Fock (RBHF) theory for the finite nuclei and achieved much better agreement with the experimental data employing only the two-body interaction~\cite{Shen2016Chin.Phys.Lett.33.102103, Shen2017Phys.Rev.C96.014316, Shen2019Prog.Part.Nucl.Phys.109.103713}, in contrast to the previous nonrelativistic Brueckner-Hartree-Fock calculations.
  Like other \textit{ab initio} calculations, the RBHF calculation is computationally consuming for heavy and even medium-mass nuclei.
  \par
  In a neutron drop, a collection of neutrons confined by an external field, for example, a harmonic trap,
only the neutron-neutron interaction exists, and the equations are easier to be solved compared with real finite nuclei.
  It thus  draws great attention~\cite{
    Pederiva2004Nucl.Phys.A742.255,
    Bogner2011Phys.Rev.C84.044306,
    Gandolfi2011Phys.Rev.Lett.106.012501,
    Maris2013Phys.Rev.C87.054318,
    Potter2014Phys.Lett.B739.445,
    Tews2016Phys.Rev.C93.024305,
    Shen2018Phys.Rev.C97.054312}
  and provides an ideal platform to link the \textit{ab initio} and DFT calculations \cite{
    Pudliner1996Phys.Rev.Lett.76.2416,
    Smerzi1997Phys.Rev.C56.2549,
    Kortelainen2014Phys.Rev.C89.054314,
    Zhao2016Phys.Rev.C94.041302R,
    Bonnard2018Phys.Rev.C98.034319,
    Shen2018Phys.Lett.B778.344,
    Ge2020Phys.Rev.C102.044304}.
  More importantly, both the single-particle energies calculated by the RBHF theory and those calculated by the DFT are quantities on the pure mean-field level, which ensures that one can make a fair comparison between the two.
  \par
  Great successes have been achieved in nuclear physics with the nuclear covariant density functional theory (CDFT)~\cite{Vretenar2005Phys.Rep.409.101,
    Meng2006Prog.Part.Nucl.Phys.57.470,
    Niksic2011Prog.Part.Nucl.Phys.66.519, Meng2016}.
  As a branch of CDFT, the RHF theory shares the common advantages of it~\cite{
    Long2006Phys.Lett.B639.242,
    Long2006Phys.Lett.B640.150,
    Long2010Phys.Rev.C81.024308,
    Liang2012Phys.Rev.C86.021302R,
    Shi2019Chin.Phys.C43.074104,
    Guo2019Chin.Phys.C43.114105,
    Yang2020Chin.Phys.C44.034102}.
  In addition, the RHF theory can take into account the tensor force via the Fock term without extra free parameters~\cite{
    Bouyssy1987Phys.Rev.C36.380,
    Long2006Phys.Lett.B640.150,
    Long2007Phys.Rev.C76.034314,
    Long2008Europhys.Lett.82.12001,
    Moreno-Torres2010Phys.Rev.C81.064327,
    Wang2013Phys.Rev.C87.047301,
    Jiang2015Phys.Rev.C91.034326,
    Zong2018Chin.Phys.C42.024101,
    Li2019Chin.Phys.C43.074107}.
  In particular, the quantitative analysis of tensor-force effects in the RHF theory was recently performed~\cite{Wang2018Phys.Rev.C98.034313}.
  According to the famous mechanism revealed by Otsuka \textit{et~al.}~\cite{Otsuka2005Phys.Rev.Lett.95.232502}, spin-orbit (SO) splittings are sensitive to  the tensor force.
  Taking the SO splittings calculated by the RBHF theory as meta-data, the strength of the tensor force was explored in the RHF theory~\cite{
    Shen2018Phys.Lett.B778.344,
    Shen2018Phys.Rev.C97.054312,
    Wang2019Phys.Rev.C100.064319,
    Zhao2020Phys.Rev.C102.034322}.
  Nevertheless, the contributions of the tensor force were not quantitatively extracted in these works.
  In the present work, we will first quantitatively verify the tensor-force effects on the SO splittings in neutron drops within the RHF theory.
  In addition, the strength of the tensor force will be further explored.
  Motivated by the idea of \textit{renormalization persistency} of the tensor force~\cite{
  Otsuka2010Phys.Rev.Lett.104.012501,
  Tsunoda2011Phys.Rev.C84.044322},
  particular attention will be paid to the density dependence of the tensor force in the nuclear medium.
  \par
  This paper is organized as follows.
  In Section~{\ref{sec:theory}}, the RHF theory and the method to evaluate the tensor-force contributions are briefly introduced.
  In Section~{\ref{sec:results}}, we clarify the tensor-force effects on the evolution of the SO splittings in neutron drops and then further explore the strength of the tensor force.
  A summary is provided in Section~{\ref{sec:summary}}.
  \section{Theoretical framework}
  \label{sec:theory}
  \par
  In the CDFT, the nucleons are considered to interact with each other by exchanging various mesons and photons \cite{Walecka1974Ann.Phys.83.491, Reinhard1989Rep.Prog.Phys.52.439, Vretenar2005Phys.Rep.409.101, Meng2006Prog.Part.Nucl.Phys.57.470, Niksic2011Prog.Part.Nucl.Phys.66.519, Meng2016, Wang2019Chin.Phys.C43.114107, An2020Chin.Phys.C44.074101, Wei2020Chin.Phys.C44.074107}. Starting from the ansatz of a standard Lagrangian density, which contains the degrees of freedom associated with the nucleon field, the meson fields, and the photon field, one can derive the corresponding Hamiltonian as
  \begin{align}
      H
  = & \,
      \int
      d^3 x \,
      \bar{\psi} \left( x \right)
      \left[
      - i \bm{\gamma} \cdot \bm{\nabla} + M
      \right]
      \psi \left( x \right)
      \notag \\
    & \,
      +
      \frac{1}{2}
      \sum_{\phi}
      \iint
      d^3 x \, d^4 y \,
      \bar{\psi} \left( x \right)
      \bar{\psi} \left( y \right)
      \Gamma_{\phi} \left( x, y \right)
      D_{\phi} \left( x, y \right)
      \notag \\
    & \,
      \qquad
      \times
      \psi \left( y \right)
      \psi \left( x \right),
      \label{Hamil}
  \end{align}
  where $\psi$ is the nucleon-field operator, and $\phi$ denotes the meson-nucleon couplings, including the Lorentz $\sigma$-scalar ($\sigma$-S), $\omega$-vector ($\omega$-V), $\rho$-vector ($\rho$-V), $\rho$-tensor ($\rho$-T), $\rho$-vector-tensor ($\rho$-VT), and $\pi$-pseudovector ($\pi$-PV) couplings, as well as the photon-vector ($A$-V) coupling.
  Here, $\Gamma_{\phi} \left(x,y\right)$ and $D_{\phi} \left(x,y\right)$ are the interaction vertex and the propagator of a given meson-nucleon coupling~$\phi$, respectively;
  their explicit expressions can be referred to Refs.~\cite{
    Bouyssy1987Phys.Rev.C36.380,
    Shi1995Phys.Rev.C52.144,
    Long2006Phys.Lett.B640.150,
    Long2007Phys.Rev.C76.034314,
    Sun2008Phys.Rev.C78.065805,
    Liang2008Phys.Rev.Lett.101.122502,
    Niu2017Phys.Rev.C95.044301,
    Wang2020arXiv:2012.13143}.
  Evidently, the photon field is not considered in the case of neutron drops.
  \par
  The nucleon-field operators, $\psi \left( x \right)$ and $\psi^{\dagger} \left( x \right)$, can be expanded on a set of creation and annihilation operators defined by a complete set of Dirac spinors
  $ \left\{ \varphi_{\alpha} \left( \bm{r} \right) \right\} $, where $\bm{r}$ denotes the spatial coordinate of $x$.
  In this work, the spherical symmetry is assumed.
  Then, the energy density functional can be obtained through the expectation value of the Hamiltonian on the trial Hartree-Fock state under the no-sea approximation~\cite{
    Walecka1974Ann.Phys.83.491}.
  Variations of the energy density functional with respect to the single-particle wave functions give the Dirac equations,
  \begin{equation}
    \int
    d \bm{r}' \,
    \hat h \left( \bm{r}, \bm{r}' \right)
    \varphi \left( \bm{r}' \right)
    =
    \varepsilon
    \varphi \left( \bm{r} \right),
  \end{equation}
  where $\hat h \left(\bm{r}, \bm{r}' \right)$ is the single-particle Hamiltonian.
  In the RHF theory, $\hat h \left(\bm{r}, \bm{r}' \right)$ contains the kinetic energy $\hat h^{\text{K}}$, the direct local potential $\hat h^{\text{D}}$, and the exchange nonlocal potential $\hat h^{\text{E}}$; see Refs.~\cite{
    Bouyssy1987Phys.Rev.C36.380,
    Long2005PhD.Thesis,
    Long2010Phys.Rev.C81.024308,
    Wang2020arXiv:2012.13143} for detailed expressions.
  Notice that the tensor force contributes only to the nonlocal potentials.
  \par
  For the RHF theory with density-dependent effective interactions, the meson-nucleon coupling strengths are taken as functions of the baryonic density $\rho_\text{b}$.
  For convenience, here we explicitly present the density dependence of the $\pi$-PV coupling, which reads
  \begin{equation}
    \label{Eq:DDPI}
    f_{\pi} \left( \rho_\text{b} \right)
    =
    f_{\pi} \left( 0 \right)
    e^{-a_{\pi} \xi},
  \end{equation}
  where $ \xi = \rho_\text{b} / \rho_{\text{sat.}} $
  with the saturation density of the nuclear matter $\rho_{\text{sat.}}$,
  and $ f_{\pi} \left( 0 \right) $ corresponds to the coupling strength at zero density.
  The coefficient $a_\pi$ determines how fast the coupling strength $f_\pi$ decreases with the increasing density.
  The density dependence of the other meson-nucleon couplings can be found in Refs.~\cite{
    Long2006Phys.Lett.B640.150,
    Long2007Phys.Rev.C76.034314}.
  \par
  The external field to keep neutron drops bound is chosen as a harmonic oscillator (HO) potential as
  \begin{equation}
    U_{\text{h.o.}} \left(r\right)=\frac{1}{2}M\omega^2r^2,
  \end{equation}
  with $\hbar\omega=10 \, \mathrm{MeV} $. It is worth noticing that the choice of the external field here is not completely arbitrary, but is optimal, as specifically discussed in Ref.~\cite{Wang2019Phys.Rev.C100.064319}.
  \par
  In Ref.~\cite{Wang2018Phys.Rev.C98.034313}, the tensor force in each meson-nucleon coupling was identified through the nonrelativistic reduction.
  They can be expressed uniformly as
  \begin{equation}
    \label{eq:Vtphi_abcd}
    \hat{\mathcal{V}}^t_{\phi}
    =
    \frac{1}{m_{\phi}^2 + \bm{q}^2}
    \mathcal{F}_{\phi}
    S_{12},
  \end{equation}
  where $m_{\phi}$ is the meson mass, $\bm{q}$ is the momentum transfer, and $S_{12}$ is the operator of the tensor force in the momentum space, which reads
  \begin{equation}
    \label{S12}
    S_{12}
    \equiv
    \left( \bm{\sigma}_1 \cdot \bm{q} \right)
    \left( \bm{\sigma}_2 \cdot \bm{q} \right)
    -
    \frac{1}{3}
    \left( \bm{\sigma}_1 \cdot \bm{\sigma}_2 \right)
    q^2.
  \end{equation}
  The coefficient $ \mathcal{F}_{\phi} $ associated with a given meson-nucleon coupling reflects the sign and the rough strength of the tensor force, as listed in Table II of Ref.~\cite{Wang2018Phys.Rev.C98.034313}.
  \par
  The method to quantitatively evaluate the contributions of the tensor force was also established in Ref.~\cite{
    Wang2018Phys.Rev.C98.034313}.
  Using this method, one can first calculate the tensor-force contributions to the two-body interaction matrix elements; the explicit formulae are referred to Appendix C in Ref.~\cite{Wang2018Phys.Rev.C98.034313}.
  Then, the contributions of the tensor force to the nonlocal potential can be obtained, and eventually, its contributions to the single-particle energies are quantitatively extracted.
  \section{Results and discussion}
  \label{sec:results}

\begin{table}
      \caption{Binding energies and radii of the neutron drops calculated by the RHF theory with the effective interaction PKO1 \cite{Long2006Phys.Lett.B640.150}, compared with the results without the tensor force (denoted by PKO1$_{\text{n.t.}}$). See the text for more details.}
      \label{Tab:bulk}
      \begin{tabular}{crrcc}
        \hline\hline
           &\multicolumn{2}{c}{$E$~(MeV)}    & \multicolumn{2}{c}{$r$~(fm)}    \\ \hline
       $N$ &    PKO1~  & PKO1$_{\text{n.t.}}$ & PKO1~   & PKO1$_{\text{n.t.}}$ \\ \hline
        4  &    60.679 &   60.518             & 2.527   & 2.526                \\  
        6  &    93.123 &   92.629             & 2.589   & 2.593                \\  
        8  &   125.681 &  125.680             & 2.703   & 2.703                \\  
        10 &   175.632 &  175.518             & 2.836   & 2.836                \\  
        12 &   223.301 &  222.924             & 2.921   & 2.923                \\  
        14 &   268.870 &  268.180             & 2.981   & 2.987                \\  
        16 &   316.018 &  315.445             & 3.067   & 3.074                \\  
        18 &   365.329 &  365.206             & 3.150   & 3.151                \\  
        20 &   414.085 &  414.084             & 3.216   & 3.216                \\  
        22 &   477.173 &  477.112             & 3.281   & 3.281                \\  
        24 &   539.286 &  539.075             & 3.336   & 3.337                \\  
        26 &   600.468 &  600.065             & 3.383   & 3.386                \\  
        28 &   660.757 &  660.163             & 3.425   & 3.430                \\  
        30 &   725.978 &  725.415             & 3.485   & 3.491                \\  
        32 &   790.552 &  790.090             & 3.538   & 3.546                \\  
        34 &   858.068 &  857.635             & 3.594   & 3.598                \\  
        36 &   925.387 &  924.758             & 3.644   & 3.646                \\  
        38 &   992.335 &  992.686             & 3.692   & 3.692                \\  
        40 &  1059.237 & 1059.236             & 3.733   & 3.733                \\  
        42 &  1137.216 & 1137.188             & 3.770   & 3.770                \\  
        44 &  1214.829 & 1214.734             & 3.805   & 3.805                \\  
        46 &  1292.080 & 1291.902             & 3.837   & 3.838                \\  
        48 &  1368.977 & 1368.722             & 3.867   & 3.869                \\  
        50 &  1445.527 & 1445.220             & 3.895   & 3.899                \\
        \hline\hline
      \end{tabular}
 \end{table}

  First, we calculate the binding energies and radii of neutron drops with the neutron number $N$ from 4 to 50 using the RHF theory with the effective interaction PKO1~\cite{Long2006Phys.Lett.B640.150}, as shown in Table~\ref{Tab:bulk}.
  {The effective interaction PKO1 is the first widely used RHF functional with density-dependent meson-nucleon coupling strengths. It has produced good descriptions for finite nuclei and nuclear matter, comparable with those of the RMF functionals.
  In particular, the tensor force is explicitly taken into account in PKO1, owing to the inclusion of the Fock terms of the relevant meson-nucleon couplings.}
  The comparison between the energies and the radii of neutron drops given by the RBHF calculation with the interaction Bonn A and those by the RHF calculation with PKO1 has been presented in Ref.~\cite{Shen2018Phys.Rev.C97.054312}.

  Here, we also calculate the energies and radii of neutron drops with PKO1 without the tensor force, i.e., the contribution of the tensor force to the nonlocal mean field is excluded in each step of the iteration before convergence is reached, using the method proposed in Ref.~\cite{Wang2018Phys.Rev.C98.034313}.
  The results are shown in Table~\ref{Tab:bulk} in comparison with those given by the full calculation with PKO1.
  It can be seen that the results with and without the tensor force are close with each other, which means that the tensor-force contributions to the energies and radii of neutron drops are negligible.
  In particular, the contributions of the tensor force almost vanish in the neutron drops with spin saturation, namely $N= 8$, $20$, and $40$.
  This is in agreement with the common understanding of the properties of the tensor force~\cite{
  Otsuka2005Phys.Rev.Lett.95.232502,
  Sagawa2014Prog.Part.Nucl.Phys.76.76}.

  The single-particle energies of neutron drops as a function of the neutron number $N$, calculated by the RBHF theory using the interaction Bonn A, are shown in Fig.~10 in Ref.~\cite{Shen2018Phys.Rev.C97.054312}.
  In general, the single-particle energies calculated by the RHF theory (which are not explicitly shown here for simplicity) reproduce the results using the RBHF theory well for states near the Fermi energy.
  However, for those far away from the Fermi energy, the differences between the results of the two models are remarkable.
  It is known that the value of single-particle energy is determined by various components of the nuclear force, such as the central force and the SO force.
  In contrast, the evolution of SO splittings is mainly determined by the tensor force and largely free from the other components~\cite{Sagawa2014Prog.Part.Nucl.Phys.76.76}.
  Thus, the evolution of SO splittings, rather than the single-particle energies themselves, should be adopted as a benchmark for the tensor force.
  As a typical representative, the relative change of SO splittings from the neutron-proton drop $^{40}$20 ($Z = 20$, $N = 20$) to $^{48}$20 one ($Z = 20$, $N = 28$) calculated by the RBHF theory using the Bonn A interaction, is employed to determine the tensor term in the Skyrme interactions~\cite{Shen2019Phys.Rev.C99.034322}.

\begin{figure}
    \includegraphics[width=0.4\textwidth]{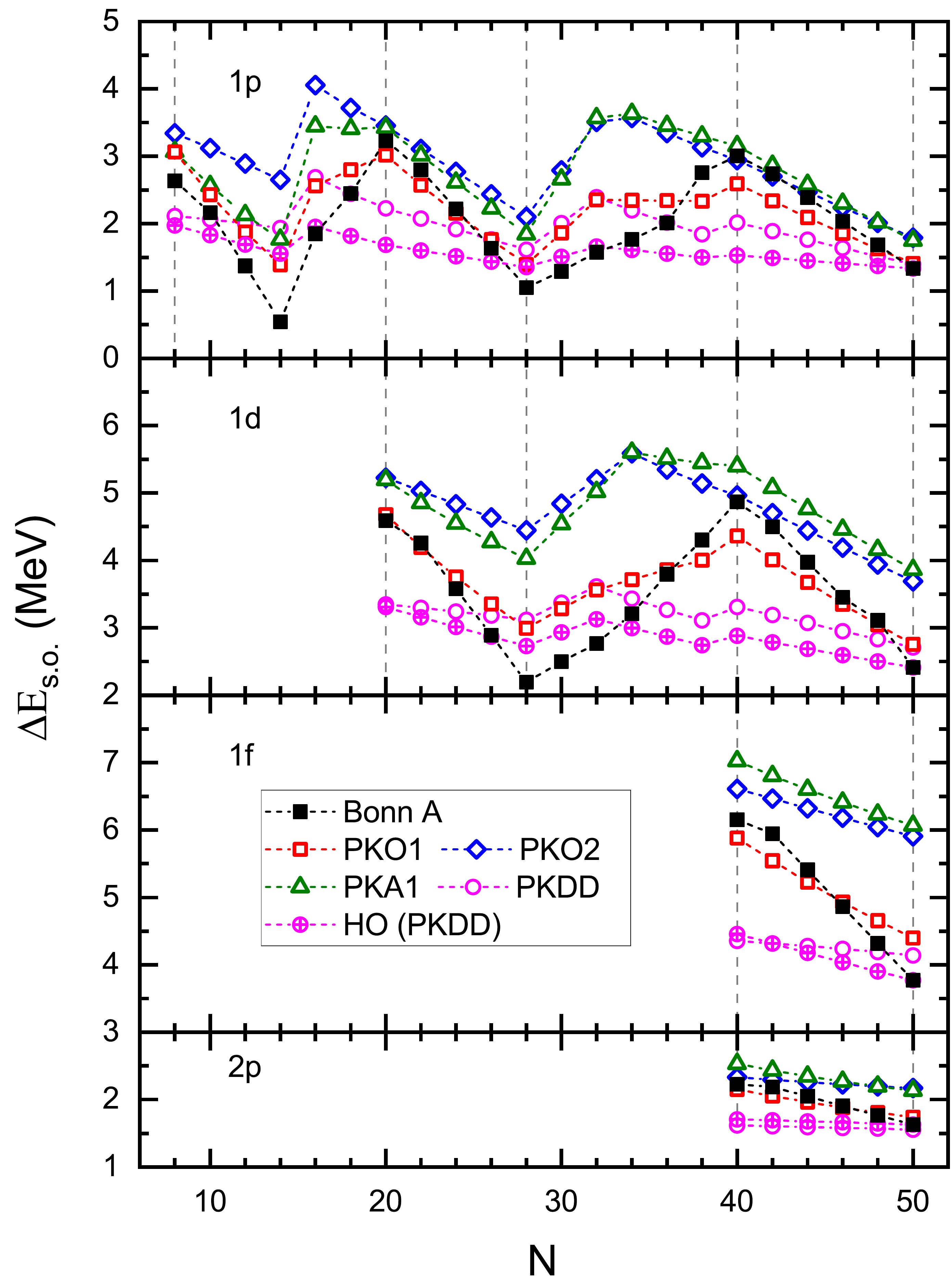}
    \caption{From top to bottom, the SO splittings of doublets $1p$, $1d$, $1f$, and $2p$ in neutron drops. The results are calculated using the RHF theory with the effective interactions PKO1~\cite{Long2006Phys.Lett.B640.150}, PKO2~\cite{Long2008Europhys.Lett.82.12001}, and PKA1~\cite{Long2007Phys.Rev.C76.034314}, as well as using the RMF theory with PKDD~\cite{Long2004Phys.Rev.C69.034319}. The results of RBHF~\cite{Shen2018Phys.Lett.B778.344} with the Bonn A interaction are shown as meta-data. The contributions of the HO potential in the RMF calculation with PKDD are also presented, denoted as ``HO (PKDD).''}
    \label{Fig:Eso}
  \end{figure}

  Displayed in Fig.~\ref{Fig:Eso} are the SO splittings of doublets $1p$, $1d$, $1f$, and $2p$ in neutron drops, calculated using both the RMF and RHF theories.
  The results of the RBHF theory obtained using the Bonn A interaction~\cite{Shen2018Phys.Lett.B778.344} are also shown for comparison, serving as the meta-data.
  One can see that the meta-data present a nontrivial pattern: the SO splittings vary monotonously and even linearly between the neighboring (sub)shells $N = 8$, $14$, $20$, $28$, $40$, and $50$.
  Such a feature arises from the characteristic spin-dependent properties of the tensor force, as pointed out in Ref.~\cite{Shen2018Phys.Lett.B778.344}.

  \par
  For the calculation with PKDD \cite{Long2004Phys.Rev.C69.034319}, which is a RMF functional with density-dependent meson-nucleon coupling strengths, the results are evidently far away from the meta-data.
  This is mainly because of the absence of the explicit tensor force in the framework of RMF~\cite{Long2008Europhys.Lett.82.12001, Shen2018Phys.Lett.B778.344}, due to the lack of the Fock term.
  As a typical representative of the RHF effective interactions, PKO1~\cite{Long2006Phys.Lett.B640.150} reproduces the pattern of meta-data qualitatively, which is attributed to the tensor force in the $\pi$-PV coupling~\cite{Shen2018Phys.Lett.B778.344}.
  Interestingly, even though the tensor force is explicitly involved in PKO2~\cite{Long2008Europhys.Lett.82.12001} and PKA1~\cite{Long2007Phys.Rev.C76.034314}, their results obviously deviate from the meta-data.
  In fact, the results of PKO2 and PKA1 are similar to that of PKDD rather than PKO1.
  To understand this phenomenon, one needs to examine the details of the tensor force arising from each meson-nucleon coupling, which are identified in Ref.~\cite{Wang2018Phys.Rev.C98.034313}.
  \par
  Among all the meson-nucleon couplings involved in the current RHF effective interactions, only the $\pi$-PV coupling produces a tensor force that matches the properties of spin dependence revealed in Ref.~\cite{Otsuka2005Phys.Rev.Lett.95.232502}, i.e., it is repulsive (attractive) when the two interacting nucleons are parallel (antiparallel) in their spin states.
  The tensor forces in the other couplings are opposite to that in the $\pi$-PV coupling in sign, as shown in Table II of Ref.~\cite{Wang2018Phys.Rev.C98.034313}.
  Meanwhile, it has been shown that the tensor force in the $\pi$-PV coupling dominates those in the other couplings.
  Thus, for PKO2, where the $\pi$-PV coupling is absent, the tensor force mainly comes from the $\omega$-V coupling.
  This means that the sign of the net tensor-force contribution in PKO2 is opposite to that in PKO1.
  That is why PKO2 cannot reproduce the pattern given by RBHF even qualitatively.
  PKA1 contains not only all the meson-nucleon couplings involved in the PKO series but also the $\rho$-T and $\rho$-VT couplings, which create tensor forces with considerable strengths.
  As mentioned above, the tensor-force contributions arising from these two couplings partially cancel the corresponding contributions from the $\pi$-PV coupling.
  Therefore, PKA1 gives worse description of the meta-data than PKO1 does.
  \par
  It is noticeable that the RMF effective interaction PKDD also presents some kinks. To elucidate where these kinks arise from, we calculate the contributions of the external HO potential, denoted as ``HO (PKDD)'' in Fig.~\ref{Fig:Eso}.
  By comparing the results of PKDD and ``HO~(PKDD),'' one can find that the kinks given by PKDD are determined by the external HO potential.
  This also explains why the results given by different RMF effective interactions are so similar to each other, as shown in Fig.~2 of Ref.~\cite{Shen2018Phys.Lett.B778.344}.
  Definitely, the external potential can also affect the shell structure given by the RHF effective interactions.
  \begin{figure}
    \includegraphics[width=0.45\textwidth]{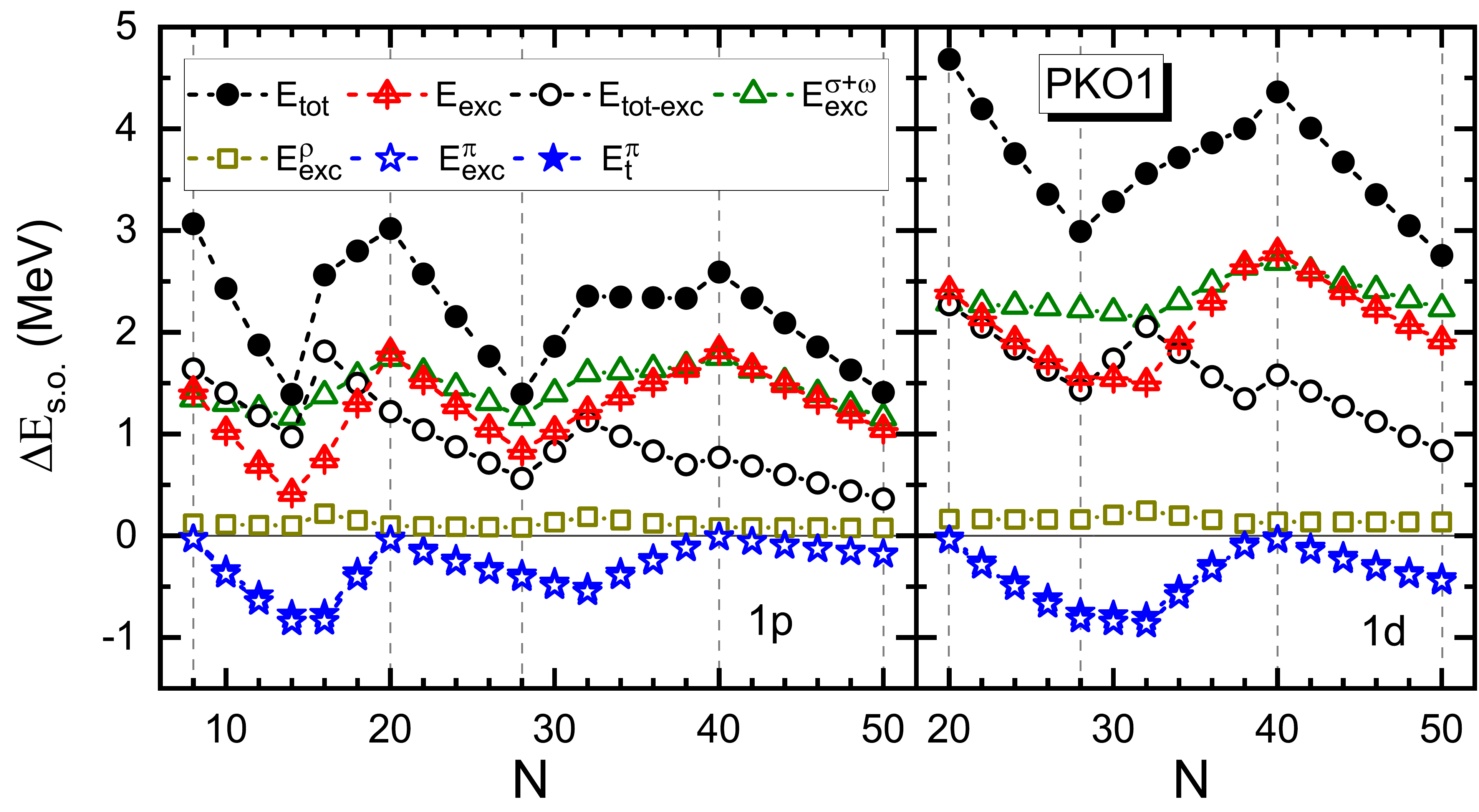}
    \caption{Contributions of the exchange terms to the SO splittings of $1p$ and $1d$ doublets in neutron drops, calculated by the RHF theory with the effective interaction PKO1. For comparison, the total SO splittings, the contributions of the tensor force from the $\pi$-PV coupling, and the results without the contribution of the Fock terms are also shown. See the text for more details.}
    \label{Fig:1p1d-exc-PKO1}
  \end{figure}
  \par
  It has been shown that the meta-data can be better reproduced by PKO1 when the coupling strength of $\pi$-PV is enhanced properly~\cite{Shen2018Phys.Lett.B778.344, Shen2018Phys.Rev.C97.054312}.
  Actually, the RHF effective interaction PKO3~\cite{Long2008Europhys.Lett.82.12001}, which contains the same kinds of meson-nucleon couplings as PKO1 but slightly stronger $\pi$-PV coupling strength, can also provide a comparable description of the meta-data.
  According to the previous works and the discussion above, it can be shown that the $\pi$-PV coupling, essentially the embedded tensor force, plays a crucial role in determining the evolutionary trend of the SO splittings in neutron drops.
  Nevertheless, all these analyses regarding the tensor-force effects are intuitive to some extent, while the quantitative investigation is missing.
  \par
  Since the tensor forces arise from only the exchange terms of the relevant meson-nucleon couplings, we first calculate the contributions of the exchange term of each meson-nucleon coupling to the SO splittings of $1p$ and $1d$ doublets. The results are shown in Fig.~\ref{Fig:1p1d-exc-PKO1}.
  It can be seen that the evolution of the SO splittings calculated by PKO1 (black filled circles) is mainly determined by the contributions of the exchange terms (red triangles with cross).
  When the contribution of the exchange terms is excluded, i.e., simply subtracted from the results of the full calculation, the trend (black open circles) becomes almost the same as that given by PKDD shown in Fig.~\ref{Fig:Eso}.
  Here, we remind that PKDD is an RMF effective interaction, which does not contain the exchange terms.
  Remarkably, the pattern given by the exchange term of the $\pi$-PV coupling (blue open stars) is almost the same as that by the total exchange term.
  The combined contributions of the exchange terms of the $\sigma$-S and $\omega$-V couplings (green open triangles) are also considerable; however, in general, they are not considerably decisive for the kinks as those of the $\pi$-PV coupling.
  The contributions of the $\rho$-V coupling are negligible because of the small coupling strength.
  \par
  It is well known that the $\pi$-PV coupling contains not only the tensor force but also central components. Thus, it is of particular significance to quantitatively evaluate the contributions of the tensor force from the $\pi$-PV coupling. We calculate the tensor-force contributions using the method developed in Ref.~\cite{Wang2018Phys.Rev.C98.034313}.
  The results are also shown in Fig.~\ref{Fig:1p1d-exc-PKO1}, denoted by the blue filled stars,  which are almost hidden behind the blue open stars. One can find that the contributions of the $\pi$-PV coupling are almost totally determined by the tensor force, whereas the role of the central force is negligible.
  In other words, the $\pi$-PV coupling affects the evolutionary trend almost fully through the embedded tensor force. We thus certify quantitatively that it is reasonable to explore the tensor-force effects by varying the $\pi$-PV coupling in previous works.
  \par
  \begin{table*}
    \begin{center}
      \caption{The rms deviations $\Delta$ (in the unit of $\mathrm{MeV}$) of the slope of the SO splittings between the neighboring (sub)shells. The upper panel gives the results of PKO1 with $f_{\pi} \left( 0 \right)$ multiplied by a factor $\lambda$ ($ \lambda > 1.0$);
        the lower panel gives the results of similar calculations but with $a_{\pi}$ multiplied by a factor $\eta$ ($ 0.0 < \eta < 1.0$).
        The optimal values of $ \lambda $ and $ \eta $ as well as the corresponding $ \Delta $ are in bold font.
        See the text for more details.}
      \label{Tab:RMS}
      \begin{tabular}{c|c|rrrrrrrr}
        \hline\hline
        \multirow{2}{*}{$f_{\pi} \rightarrow \lambda f_{\pi}$}
        & $\lambda$ & \multicolumn{1}{c}{$1.0$} & \multicolumn{1}{c}{$1.1$} & \multicolumn{1}{c}{$1.2$} & \multicolumn{1}{c}{$1.3$} & \multicolumn{1}{c}{$1.4$} & \multicolumn{1}{c}{$\mathbf{1.42}$} & \multicolumn{1}{c}{$1.5$} & \multicolumn{1}{c}{$1.6$} \\ \cline{2-10}
        & $\Delta$ & $ 0.0909 $ & $ 0.0758 $ & $ 0.0601 $ & $ 0.0456 $ & $ 0.0366 $ & $ \mathbf{0.0362} $ & $ 0.0402 $ & $ 0.0561 $ \\
        \hline
        \multirow{2}{*}{$a_{\pi} \rightarrow \eta a_{\pi}$}
        & $\eta$ & \multicolumn{1}{c}{$0.7$} & \multicolumn{1}{c}{$0.6$} & \multicolumn{1}{c}{$0.5$} & \multicolumn{1}{c}{$0.4$} & \multicolumn{1}{c}{$\mathbf{0.33}$} & \multicolumn{1}{c}{$0.3$} & \multicolumn{1}{c}{$0.2$} & \multicolumn{1}{c}{$0.1$} \\ \cline{2-10}
        & $\Delta$ &  $ 0.0645 $ & $ 0.0543 $ & $ 0.0439 $ & $ 0.0356 $ & $ \mathbf{0.0334} $ & $ 0.0341 $ & $ 0.0441 $ & $ 0.0643 $ \\
        \hline\hline
      \end{tabular}
    \end{center}
  \end{table*}
  \par
  It is notable that, in both the RHF and RBHF calculations, the filling approximation is adopted, i.e., the last occupied levels are partially occupied with equal probabilities for the degenerate states.
  Such an approximation may affect the levels which are not fully occupied but does not affect the neutron drops with closed (sub)shells.
  If we consider only the neutron drops with closed (sub)shells, i.e., those with $N = 8$, $14$, $20$, $28$, $40$, and $50$, we can avoid the distraction from the filling approximation.
  Here, we define the slope of the SO splittings in neutron drops with respect to the neutron numbers as
  \begin{equation}
    \label{Eq:L_i}
    L_{\text{$N_1$-$N_2$}} = \frac{\Delta E_{\text{s.o.}} \left( N_2 \right) - \Delta E_{\text{s.o.}} \left( N_1 \right)}{N_2 - N_1},
  \end{equation}
  where $\Delta E_{\text{s.o.}} \left(N\right)$ is the SO splitting in the neutron drop with the neutron number $N$.
  For the neutron drops with closed (sub)shells, the combination of ($N_1$-$N_2$) has the following choices:
  ($8$-$14$), ($14$-$20$), ($20$-$28$), ($28$-$40$), and ($40$-$50$).
  Following the strategy proposed in Ref.~\cite{Shen2018Phys.Lett.B778.344}, i.e., multiplying the $f_{\pi} \left( 0 \right)$ in PKO1 by a factor $\lambda $ ($\lambda > 1.0$),
  we calculate once again the SO splittings in neutron drops.
  Then, we obtain the slopes defined above for different SO doublets and the root-mean-square (rms) deviations (denoted by $\Delta$) with respect to the RBHF results
  as
  \begin{equation}
    \Delta=\sqrt{\frac{\sum_i \left(L_i^{\text{RHF}}-L_i^{\text{RBHF}}\right)^2}{10}},
  \end{equation}
  where $L_i^{\text{RHF}}$ ($L_i^{\text{RBHF}}$) is the slope calculated by the RHF (RBHF) theory, and $i$ runs over all the possible combinations of ($N_1$-$N_2$) for different SO doublets.
  The results are presented in the upper half of Table~\ref{Tab:RMS}.
  It can be seen that $\lambda=1.42$ gives the smallest $\Delta$ among the values of $\lambda$, which is $0.0362 \, \mathrm{MeV}$.
  This, in general, agrees with the conclusion in Ref.~\cite{Shen2018Phys.Lett.B778.344}.
  \par
\begin{figure}
    \includegraphics[width=0.4\textwidth]{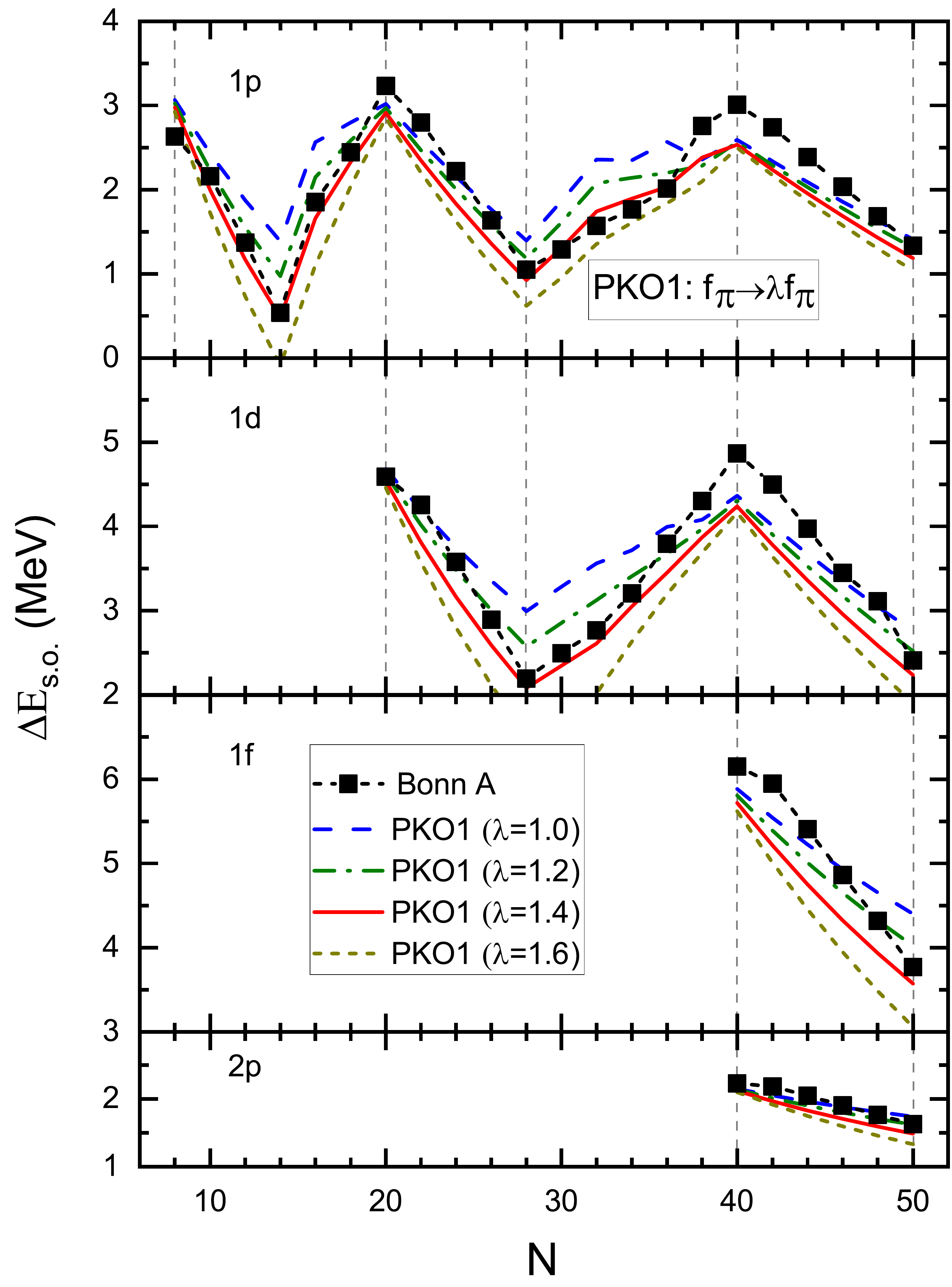}
    \caption{From top to bottom, the SO splittings of doublets $1p$, $1d$, $1f$, and $2p$ in neutron drops. The calculations are performed by the RHF theory with the effective interaction PKO1, of which the $\pi$-PV coupling strength $f_{\pi}(0)$ is multiplied by a factor $\lambda$ ($\lambda = 1.0$, $1.2$, $1.4$, and $1.6$). The results of RBHF obtained using the Bonn A interaction are also shown for comparison.}
    \label{Fig:Ndrop-PKO1-xfpi}
  \end{figure}

  To present a clearer comparison, we consider the results of $\lambda = 1.0$, $1.2$, $1.4$, and $1.6$ as examples and display them in Fig.~\ref{Fig:Ndrop-PKO1-xfpi}.
  One can find that the results of $\lambda = 1.2$ are slightly closer to the meta-data compared with the original PKO1.
  The meta-data can be reproduced better when $ \lambda = 1.4 $,
  which is quite close to the optimal value of $ 1.42 $.
  If $\lambda$ becomes significantly larger, the results get worse in general.
  Accordingly, it can be seen that the results of $ \lambda =1.6 $ evidently deviate from the meta-data.

  \par
  In addition to the strength of the tensor force, the properties of the tensor force in nuclear medium is also of great interests and still under discussion~\cite{Sagawa2014Prog.Part.Nucl.Phys.76.76}.
  It has been argued that the bare tensor force does not undergo significant renormalization in the medium, which is denoted as the \textit{tensor renormalization persistency}~\cite{Otsuka2010Phys.Rev.Lett.104.012501, Tsunoda2011Phys.Rev.C84.044322}.
  In other words, the effective tensor force would be similar to the bare one.
  If this is true, one could also anticipate to verify such a property in the framework of DFT.
  Within the RHF theory with density-dependent effective interactions, the renormalization can be, to a large extent, reflected by the density dependence of the coupling strengths.
  Minor renormalization can naturally manifest as weak density dependence.
  For the $\pi$-PV coupling, which serves as the main carrier of the tensor force, \textit{tensor renormalization persistency} requires that the coefficient of density dependence $a_{\pi}$ should be small, as indicated by Eq.~\eqref{Eq:DDPI}.
  In our previous work~\cite{Wang2020arXiv:2012.13143}, we have shown that weakening the density dependence of the $\pi$-PV coupling, i.e., reducing $a_{\pi}$, can improve the description of the shell-structure evolution in the $ N = 82$ isotones and the $Z = 50$ isotopes more efficiently, compared with enlarging $f_{\pi}(0)$ with a factor.
  Inevitably, the beyond-mean-field correlations in the experimental single-particle energies make the comparison with the results on the mean-field level ambiguous.
  With the meta-data given by the RBHF theory, which are also on the pure mean-field level, we can further explore the \textit{tensor renormalization persistency} in a more convincing way.
    \par
  \begin{figure}
    \includegraphics[width=0.4\textwidth]{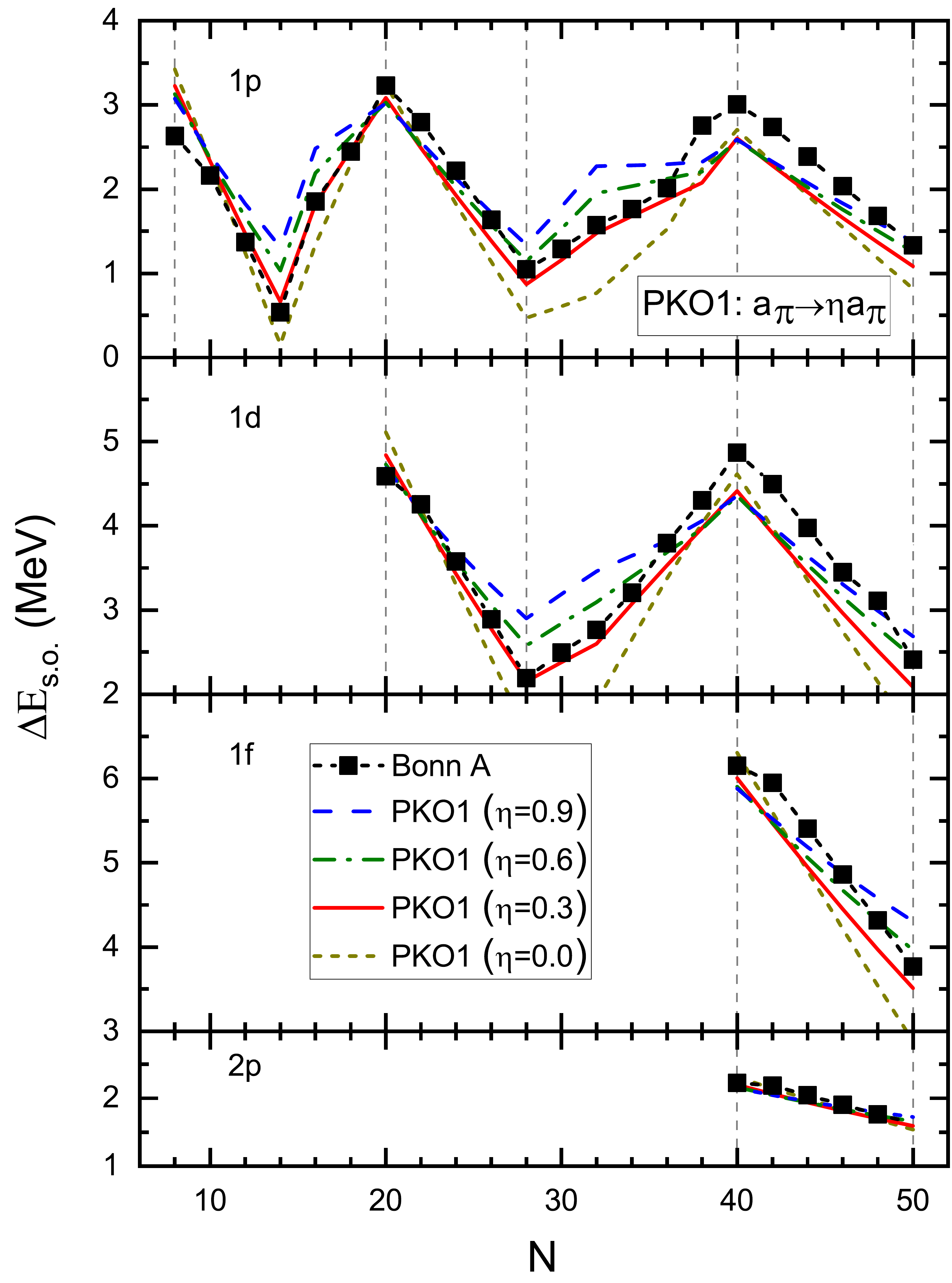}
    \caption{Similar to Fig.~\ref{Fig:Ndrop-PKO1-xfpi}, but $a_{\pi}$ is multiplied by a factor $\eta$ ($\eta = 0.9$, $0.6$, $0.3$, and $0.0$).}
    \label{Fig:Ndrop-PKO1-xapi}
  \end{figure}
  \par
  Aiming at this goal, we recalculate the SO splittings in neutron drops with the RHF theory using PKO1, but the coefficient of density dependence $a_{\pi}$ is multiplied by a factor $\eta$
  ($ 0.0 < \eta < 1.0$).
  We present in the lower half of Table~\ref{Tab:RMS} the rms deviations of the slope of the SO splittings between the neighboring (sub)shells.
  One can see that the smallest deviation is obtained when $\eta = 0.33$, which is $0.0334 \, \mathrm{MeV}$.
  We also find that the minimum value of $\Delta$ for modification with $\eta$ is smaller than that with $\lambda$,
  which means that the modification with $\eta$ is more adequate.
  Thus, one can conclude that weakening the density dependence is an available and efficient way to improve the description of the evolution of SO splittings.
  It should be stressed that this conclusion is reached by the comparison between the CDFT calculation and \textit{ab initio} one, both of which belong to the pure mean-field level.
  Even though we did not perform a complete refitting procedure yet, we indeed provided a semiquantitative support for the \textit{renormalization persistency} of the tensor force.
  \par
  To illustrate the discussion above more clearly, the results with $\eta= 0.0$, $0.3$, $0.6$, and $0.9$ are considered as examples and shown in Fig.~\ref{Fig:Ndrop-PKO1-xapi}, in comparison with the RBHF results.
  It can be seen that a smaller $a_{\pi}$, which means weaker density dependence, gives a better description of the evolution of SO splittings.
    When $\eta=0.3$, which is quite close to the optimal value of $0.33$, the meta-data are reproduced quite well.
  If $\eta$ is too small, the results become visibly worse. Eventually, when $\eta =0.0$, the deviation from the meta-data is much more significant.

  \section{Summary}
  \label{sec:summary}
  \par
  In this work, we have investigated the tensor-force effects on the evolution of SO splittings in neutron drops within the framework of the RHF theory.
  The corresponding results of the RBHF calculation with the Bonn A interaction were adopted as meta-data. Since the results of the RHF theory and the meta-data calculated by the RBHF theory are both within the pure mean-field level, fair comparisons can be made between them.
  Through a qualitative analysis of the results using the RHF effective interactions PKO1, PKO2, and PKA1, we confirmed that the tensor force in the effective interactions plays a crucial role in reproducing the meta-data.
  Meanwhile, for the RMF effective interactions, it was found that the evolution of SO splittings is mainly determined by the external HO potential.
  Moreover, we found that the contributions from the exchange terms almost fully determine the evolution of SO splittings.
  Among all the meson-nucleon couplings, the exchange term of $\pi$-PV coupling plays the dominant role.
  In particular, the tensor-force contribution was extracted, and it was found that the $\pi$-PV coupling affects the evolutionary trend through the embedded tensor force, while its central force has almost invisible effects.
  This conclusion verifies quantitatively that it is reasonable to explore the tensor-force effects by varying the $\pi$-PV coupling strength.
  In other words, the evolution of SO splittings belongs to the observables that can constrain the strength of the tensor force.
  \par
  The strength of the tensor force was explored by enlarging $f_{\pi}$ in two different ways: (i) multiplying a factor $\lambda$ ($\lambda > 1.0$) as a whole and (ii) weakening the density dependence by multiplying a factor $\eta$ ($0.0 < \eta < 1.0$) for the coefficient $a_{\pi}$. To avoid possible distractions from the filling approximation adopted in the RHF and RBHF calculations, we took into account only the neutron drops with (sub)shell closure and calculated the slopes of the SO splittings between the neighboring (sub)shells. Judging from the rms deviations of the selected slopes, with respect to those calculated from the meta-data, we found that when $\lambda \simeq 1.4$ or $\eta \simeq 0.3$, the meta-data are reproduced best. In particular, weakening the density dependence appears to be slightly better than enlarging $f_{\pi}$ with a factor. In this manner, we provide a semiquantitative support for the \textit{renormalization persistency} of the tensor force. Naturally, in practice, an overall refitting of the parameters in the RHF effective interaction is necessary. Work in this direction is in progress.

\vspace{10mm}%


\end{document}